\documentclass[preprintnumbers,amsmath,amssymb,prd]{revtex4}

\usepackage{graphicx}
\usepackage{dcolumn}
\usepackage{bm}

\newcommand{\ihmpc}{\, h\, {\rm Mpc}^{-1}}

\newcommand{\lyaf}{Ly$\alpha$ forest}

\newcommand{\ordertwo}{{\mathcal{O}(k_c^{-2})}}
\newcommand{\vk}{{\mathbf k}}
\newcommand{\vx}{{\mathbf x}}
\newcommand{\vs}{{\mathbf s}}

\begin{document}

\title{Gravitational redshift and other redshift-space distortions of the 
imaginary part of the power spectrum}

\author{Patrick McDonald}
\email{pmcdonal@cita.utoronto.ca}
\affiliation{Canadian Institute for Theoretical Astrophysics, University of
Toronto, Toronto, ON M5S 3H8, Canada}

\date{\today}

\begin{abstract}

I extend the usual linear-theory formula for large-scale clustering in 
redshift-space to include gravitational redshift. The extra contribution to the
standard galaxy power spectrum is suppressed by $k_c^{-2}$, where $k_c=c k/a H$
($k$ is the wavevector, $a$ the expansion factor, and $H=\dot{a}/a$), and is 
thus effectively limited to the few largest-scale modes and very difficult to 
detect; however, a correlation, $\propto k_c^{-1}$, is generated between the 
real and imaginary parts of the Fourier space density fields of two different 
types of galaxy, which would otherwise be zero, i.e., the cross-power spectrum 
has an imaginary part: $P_{ab}(k,\mu)/P(k)=\left(b_a+f\mu^2\right)
\left(b_b+f\mu^2\right)-i\frac{3}{2}\Omega_m\frac{\mu}{k_c}\left(b_a-b_b\right)
+\mathcal{O}(k_c^{-2})$, where $P(k)$ is the real-space mass-density power 
spectrum, $b_i$ are the galaxy biases, $\mu$ is the cosine of the angle between
the wavevector and line of sight, and $f=d\ln D/d\ln a$ ($D$ is the linear 
growth factor). The total signal-to-noise of measurements of this effect is not
dominated by the largest scales -- it converges at $k\sim 0.05\ihmpc$. This
gravitational redshift result is pedagogically interesting, but naive in that 
it is gauge dependent and there are other effects of similar form and size, 
related to the transformation between observable and proper coordinates. I 
include these effects, which add other contributions to the coefficient of 
$\mu/k_c$, and add a $\mu^3/k_c$ term, but don't qualitatively change the 
picture. The leading source of noise in the measurement is galaxy shot-noise, 
not sample variance, so developments that allow higher S/N surveys can make 
this measurement powerful, although it would otherwise be only marginally 
detectable in a JDEM-scale survey.

\end{abstract}


\maketitle

\section{Introduction}

Redshift-space distortions by peculiar velocities were first computed
in \cite{1987MNRAS.227....1K}, leading to the well-known formula for the galaxy
power spectrum, 
$P_g(k,\mu)=\left(b+f \mu^2\right)^2 P(k)$ where $b$ is the bias, 
$f=d\ln D/d\ln a$, $D$ is the linear growth factor, $a$ the expansion 
factor,
$\mu$ is the cosine of the angle between the wavevector $\vk$ and the line of 
sight, and $P(k)$ is the real space mass-density power spectrum. 
Measurements of these distortions can provide powerful constraints on 
cosmology, isolating the bias-independent quantity $f^2 P(k)$, including it's
redshift evolution \cite{2007PhRvL..99n1302Z,2008arXiv0810.0323M,
2009MNRAS.tmp..925W,2008arXiv0802.2416Z,2009MNRAS.393..297P}. It has
always been known that the formula of \cite{1987MNRAS.227....1K} is approximate
in a variety of ways, which must be improved as precision of measurements 
improves. The most 
important corrections are for non-linearity, which are important on relatively
small scales where we have the most statistical power
\cite{1998MNRAS.301..797H,2004PhRvD..70h3007S,2007PASJ...59.1049N,
2008PhRvD..77f3530M,2008PhRvD..78h3519M,
2008PhRvD..78j3512S,2009arXiv0906.0507T}.
The form of redshift-space distortions can be modified even on linear scales
by non-linear transformations of the already distorted field 
\cite{2000ApJ...543....1M,2003ApJ...585...34M,2009arXiv0902.0991M}
or possibly selection effects \cite{2009arXiv0903.4929H}.
Finally, on large scales there are relativistic effects
\cite{2009PhRvD..79b3517Y,2009arXiv0907.0707Y}, 
which will be the subject of this paper.

We normally assume that the Universe is intrinsically homogeneous and isotropic.
Our line of sight breaks this symmetry when we perform a redshift survey, 
however, standard redshift-space distortions due to peculiar velocities do not
break reflection symmetry along the line of sight, i.e., the survey would 
look the same if viewed from the opposite direction (ignoring effects on the 
photons as they travel to us, and anything else not included in the standard
result of \cite{1987MNRAS.227....1K}). 
Gravitational redshifts do break this symmetry, 
e.g., the redshift of a photon coming out of a potential well is independent of
viewing angle, so the apparent displacement is always away from the observer,
i.e., it changes sign if the viewer moves to the opposite side.
Note, however, that when measuring a standard auto-correlation function
$\xi\left(r_\parallel =x_\parallel-x_\parallel^\prime\right)=
\left<\delta(x_\parallel)\delta(x_\parallel^\prime)\right>$ (where 
$x_\parallel$ is the radial coordinate and I have suppressed the transverse
coordinates because they are irrelevant to this discussion),
there is nothing to distinguish $\xi(r_\parallel)$ from $\xi(-r_\parallel)$, 
i.e., if the galaxies are identical, there is no way to detect an exchange of
their positions. If we consider the cross-correlation of two types of galaxies,
$a$ and $b$, the situation changes -- there can in principle be a difference
between $\left<\delta_a(x_\parallel)\delta_b(x_\parallel^\prime)\right>$ and
$\left<\delta_b(x_\parallel)\delta_a(x_\parallel^\prime)\right>$, i.e., 
equivalently, between $\xi_{ab}(r_\parallel)$ and $\xi_{ab}(-r_\parallel)$.
The power spectrum, the Fourier transform of the correlation function, is 
normally real because the correlation function is even in $r_\parallel$. If the 
correlation function has a component that is odd in $r_\parallel$, the 
power spectrum has an imaginary part. Such an imaginary part, generated by 
gravitational redshift and other redshift-space distortions not included in
\cite{1987MNRAS.227....1K}, will be the 
subject of this paper.  

The plan of the paper is as follows:
In \S\ref{secnaive} I compute the effect of gravitational redshift alone.
This calculation is intended to be pedagogical, presenting the basic form of
the new effect in a calculation that is easy to understand -- it does not 
contain all of the terms of similar magnitude and is not gauge invariant.
In \S\ref{secYoo} I add the additional terms recently 
computed by \cite{2009arXiv0907.0707Y}, and estimate how precisely it will be
possible to measure this effect
in the future. Finally, in \S\ref{secconclusions} I briefly summarize and
discuss the results. 

\section{Naive gravitational redshift effect \label{secnaive}}

I first compute the redshift-space distortion due to gravitational redshift in
the way that standard velocity-induced distortions have been computed
\cite{1987MNRAS.227....1K}. It turns
out that other terms are similarly important, but this calculation contains all
of the qualitatively new $\vk$ dependence (where $\vk$ is the wavevector).
I am going to compute what for the CMB would be called the primary 
fluctuations, dependent only on the potential at the location of the galaxy,
not on an integral along the photon's path to us.

In the small-separation limit, the radial coordinate 
measured by a redshift survey is
\begin{equation}
\frac{c \Delta \lambda}{\lambda}=H a \Delta x_\parallel
+\Delta v_\parallel-\frac{\Delta \psi}{c}
\end{equation}
where the first term is the usual Hubble expansion of the Universe 
($H=\dot{a}/a$), the 
second is the Doppler shift due to radial peculiar velocity $v_\parallel$, 
and the last is gravitational redshift due to potential $\psi$.
The gravitational redshift is usually ignored, for what we will see has been 
a good reason, for relatively small-volume existing surveys.
The apparent comoving distance is
\begin{equation}
\Delta s_\parallel=\Delta x_\parallel
+\left(H a\right)^{-1}\Delta v_\parallel-
\left(c H a\right)^{-1} \Delta \psi ~.
\end{equation}
I assume that the conversion from transverse angular separation to 
Hubble velocity separation is known, so 
the components of $\vx$ and $\vs$ transverse to the line of sight are 
equivalent.  Including the uncertainty in this angle-velocity conversion leads 
to the Alcock-Paczynski test. In addition to assuming that we know the 
functions $H(z)$ and $D_A(z)$, I also assume that we know at what 
redshift to evaluate them -- of course, for a real survey we do not, we only 
know the observed
redshift of the galaxies, not the expansion factor in the background Universe.
This, and similar issues, will lead to the additional terms, computed by 
\cite{2009arXiv0907.0707Y}, which I add in the next section.
Density in redshift space, $\rho_s$ is related to density in real space,
$\rho$, by $\rho_s d^3\vs = \rho d^3\vx$.  Using
\begin{equation}
\frac{\partial s_\parallel}{\partial x_\parallel}=
1+\left(H a\right)^{-1}\frac{\partial v_\parallel}{\partial x_\parallel}-
\left(c H a\right)^{-1} \frac{\partial \psi}{\partial x_\parallel}~,
\end{equation}
and making the distant observer approximation \cite{2008MNRAS.389..292P}, 
gives
\begin{equation}
\left(1+\delta_s \right)\left[
1+\left(H a\right)^{-1}\frac{\partial v_\parallel}{\partial x_\parallel}-
\left(c H a\right)^{-1} \frac{\partial \psi}{\partial x_\parallel} \right]
=\left(1+b~\delta \right)~,
\end{equation}
or
\begin{equation}
\delta_s=b~\delta -
\left(H a\right)^{-1}\frac{\partial v_\parallel}{\partial x_\parallel}+
\left(c H a\right)^{-1} \frac{\partial \psi}{\partial x_\parallel}~, 
\end{equation}
where I am working to linear order in the perturbations.
In Fourier space this is 
\begin{equation}
\delta_{s \vk}=b~\delta_\vk + i 
\left(H a\right)^{-1} k_\parallel v_{\parallel \vk}- i 
\left(c H a\right)^{-1} k_\parallel \psi_\vk ~.
\end{equation}
The velocity is given in linear theory by
\begin{equation}
v^\parallel_\vk=-i f \frac{\mu}{ k_c} \delta_\vk ~, 
\end{equation}
where $\mu=k_\parallel/k$ and
$k_c=c~k/a~H$ is the wavevector in the observable Hubble velocity units over 
the speed of light ($v$ is also normalized by the speed of light). 
The potential is
\begin{equation}
\psi_\vk=-\frac{3}{2}\Omega_m\left(a\right) k_c^{-2} \delta_\vk~.
\end{equation}
Then,
\begin{equation}
\label{eqmaindensity}
\delta_{s \vk} = 
\left[b + f \mu^2  + 
i \frac{3}{2} \Omega_m \frac{\mu}{k_c} \right]\delta_\vk ~.
\end{equation}
$\Omega_m=\Omega_m\left(a\right)$ here, and in general
such quantities should be assumed to be time dependent, rather than the $z=0$
value, unless otherwise indicated. 
The power spectrum of $\delta_{s \vk}$ is
\begin{equation}
\left<\delta_{s \vk} \delta_{s \vk^\prime} \right>=
\left(2 \pi\right)^3 \delta^{\rm D}\left(\vk+\vk^\prime\right)
\left[\left(b+f \mu^2\right)^2+
\left(\frac{3}{2}\Omega_m \frac{\mu}{k_c}\right)^2
\right] P\left(k\right)~, 
\end{equation}
where $P(k)$ is the real space mass density power spectrum.
The fact that the new term is proportional to $(c k/ a H)^{-2}$ means that it
is only significant on very large scales, if at all.
It is useful to plug in some numbers to see just how bad
this is. A 100 cubic Gpc/h JDEM-scale survey \cite{2009arXiv0901.0721A} 
($\sim 2/3$ of the sky over $1<z<2$) has
a minimum $k\simeq 2 \pi/V^{1/3} =0.0014 \ihmpc$ or $k_c\simeq 4.5$, i.e., 
one should not be tempted to think that $k_c\simeq 1$ is reachable (every bit
of volume out to $z\sim 5$ would only reach $k_c\sim 2$). This means that, 
for the 100 cubic Gpc/h survey, for the best case $\mu=1$, the quantity 
$\left(\frac{3}{2}\Omega_m \frac{\mu}{k_c}\right)^2=0.08$ for the largest 
(fully sampled) modes in the survey. The standard power, which sets the noise 
level,
has $\left(b+f \mu^2\right)^2=3.7$, for a modest $b=1$, and $f(z=1.5)=0.92$, 
i.e., 45 times larger
than the gravitational redshift effect, which clearly cannot be detected in 
the straightforward auto-power spectrum of galaxies (the
$k^{-2}$ decline in the signal means that one can not sum many modes to 
overcome the small size of the effect).
On the bright side, if one is interested in detecting 
non-Gaussianity in the LSS power spectrum \cite{2008PhRvD..77l3514D,
2008PhRvD..78l3519M,2008PhRvD..78l3507A,2008PhRvD..78l3534T,
2008arXiv0811.4176P,2008ApJ...684L...1C,
2009JCAP...03..004S,2009MNRAS.396...85D,2009arXiv0902.2013G,
2009MNRAS.394..133C,
2009arXiv0904.0497J,2009arXiv0904.4257G,2009arXiv0905.0717S,
2009arXiv0907.2257D}, the 
smallness of this kind of effect is a good thing (there will be other, 
isotropic, terms of similar order \cite{2009arXiv0907.0707Y}, which would look 
like the effect of the local model of non-Gaussianity).

We would really like to find a cross-term between the gravitational 
redshift term in Eq. (\ref{eqmaindensity}) and the standard terms, which would 
be proportional to $k^{-1}$ instead of $k^{-2}$, 
and we would also like to somehow avoid competing with the 
standard power as background noise.  
It turns out that if we cross-correlate two different types of fields with
different bias we can satisfy both of these desires.
\begin{equation}
\left<\delta^a_{s \vk} \delta^b_{s \vk^\prime} \right>=
\left(2 \pi\right)^3 \delta^{\rm D}\left(\vk+\vk^\prime\right)
\left[\left(b_a+f \mu^2\right)\left(b_b+f \mu^2\right)+
\left(\frac{3}{2}\Omega_m \frac{\mu}{k_c}\right)^2
-i\frac{3}{2}\Omega_m\frac{\mu}{k_c} \left(b_a-b_b\right)\right] 
P\left(k\right)~.
\end{equation}

Note that one can isolate the imaginary term operationally by computing
\begin{equation}
P_I^{ab}(\vk) \equiv
\left<{\rm Im}\left[\delta_{s\vk}^a \delta_{s\vk}^{b\star}\right]\right> =
\left<\delta_{s\vk}^{a I} \delta_{s\vk}^{b R}-
\delta_{s\vk}^{a R} 
\delta_{s\vk}^{b I}\right>  ~,
\end{equation}
where $\delta_{s\vk}^{i R}$ and $\delta_{s\vk}^{i I}$ are the real and 
imaginary parts of the galaxy density field.
This is directly computable from observations, without knowing anything about
the underlying theory. We generally would not compute it because it is 
assumed to be zero.

It may not be completely obvious how to compute the expected error on a 
measurement of $P_I^{ab}(\vk)$, and, for the next section, it will be useful
to go through the whole calculation for a 
galaxy density field of the following general form:
\begin{equation}
\delta_{s\vk}^i =(R^i_\vk +i I^i_\vk)(\delta_\vk^R+i \delta_\vk^I)
\end{equation}
where $\delta_\vk^R$ and $\delta_\vk^I$ are the real and imaginary parts of the
mass density field and $R^i_\vk$ and $I^i_\vk$ are the real and imaginary 
parts of the 
bias-type coefficients, for galaxy type $i$ 
(i.e., $I$ would traditionally be zero, while in the 
calculation of this section $I_\vk = \frac{3}{2}\Omega_m \frac{\mu}{k_c}$).
Then
\begin{equation}
\label{eqgeneralPI}
\left<{\rm Im}\left[\delta_{s\vk}^a \delta_{s\vk}^{b\star}\right]\right> 
=\left(I^a_\vk R^b_\vk
 -R^a_\vk I^b_\vk\right)\left<\delta_\vk^{I 2}+
 \delta_\vk^{R 2}\right> = \left(I^a_\vk R^b_\vk-R^a_\vk I^b_\vk\right)P(k)~.
\end{equation}
A key fact about this calculation is that, if $I^i=0$,
${\rm Im}\left[\delta_{s\vk}^a \delta_{s\vk}^{b\star}\right]=0$
{\it mode-by-mode}, i.e., before taking any expectation
value. This means that the sample variance will also go to zero
if $I\rightarrow 0$.
Note that I have been ignoring shot-noise. If I continue to ignore 
it, I find the variance
\begin{equation}
\left<\left({\rm Im}\left[\delta_{s\vk}^a \delta_{s\vk}^{b\star}\right]-
\left<{\rm Im}\left[\delta_{s\vk}^a \delta_{s\vk}^{b\star}\right]\right>
\right)^2\right>=\left(I^a_\vk R^b_\vk
 -R^a_\vk I^b_\vk\right)^2 P(k)^2 =
\left<{\rm Im}\left[\delta_{s\vk}^a \delta_{s\vk}^{b\star}\right]\right>^2~,
\end{equation}
i.e., the error on
$\left<{\rm Im}\left[\delta_{s\vk}^a \delta_{s\vk}^{b\star}\right]\right>$
follows the usual rule for a power spectrum, with $S/N=1$ per mode.
Remarkably, this shows that the fractional errors on the imaginary part 
of the power spectrum in the cosmic variance limit are just as small as the 
fractional errors on the standard power spectrum, in spite of the fact that
the real part is much larger and one might have expected it to provide 
background noise. This feature relies on the fact that the fields for the two
types of galaxy trace the same underlying fluctuations. Once noise is added, 
the situation is not quite as rosy, although still quite good.
I assume the noise is the standard Poisson sampling noise, uncorrelated 
between the two types of galaxy. I write the measured density as 
$\tilde{\delta}_{s\vk}^i= \delta_{s\vk}^i+\epsilon_\vk^i$. The noise makes no 
mean contribution to the imaginary part of the cross-power spectrum, but it 
does add variance in the measurement:
\begin{equation}
\left<\left({\rm Im}\left[\tilde{\delta}_{s\vk}^a 
\tilde{\delta}_{s\vk}^{b\star}\right]-P_I^{ab}(\vk)\right)^2\right>=
P_I^{ab}(\vk)^2+\frac{1}{2}N^a P^b(\vk)+\frac{1}{2}N^b P^a(\vk)+
\frac{1}{2}N^a N^b ~,
\end{equation}
where $N^i$ is the noise power for galaxy type $i$, and $P^i(\vk)$ is the 
auto-power spectrum for galaxy type $i$ (always real). In the relevant regime, 
the errors here will be dominated by the terms containing $P^i\propto R^2$. 
Remember, however, that the imaginary signal itself is proportional to $R I$, 
so, at an order of magnitude level (assuming the biases do not nearly cancel), 
the $R$'s cancel in the $S/N$ ratio, leaving 
$(S/N)^2\propto I^2 P/N$, i.e., the detectability of the signal depends on the 
size of the gravitational redshift effect compared to the noise, not compared
to the real part of the signal. 
As the noise is reduced, the measurement improves, until one reaches 
the limit where $S/N=1$ per mode (in practice, this limit will be very 
difficult to reach).  More concretely,
\begin{equation}
\left(\frac{S}{N}\right)^2_\vk \simeq 
\frac{2 \left(I^a_\vk R^b_\vk -R^a_\vk I^b_\vk\right)^2 P(k)}
{R^{b 2}_\vk N^a+R^{a 2}_\vk N^b}~.
\end{equation}  

The signal-to-noise of a power spectrum measurement aimed at a simple detection
of some effect in a survey with volume $V$ is roughly 
\begin{equation}
\label{eqSN}
\left(\frac{S}{N}\right)^2=
\frac{V }{8 \pi^2} \int_{k_{\rm min}}^{k_{\rm max}} k^2 dk  \int_{-1}^1 d\mu
\left(\frac{S}{N}\right)^2\left(k,\mu\right)~.
\end{equation}
If $(S/N)^2(k)\propto k^\gamma$ with 
$\gamma<-3$, the integral converges quickly at the low $k$ end, 
while if $\gamma>-3$
the total $S/N$ continues to increase with increasing $k_{\rm max}$.
For $S/N(k)$ rolling with $k$, the integral will be dominated by the scale 
where
$\gamma\simeq -3$. In the case at hand, $\gamma(k)=n_{\rm eff}(k)-2$, where
$n_{\rm eff}(k)=d\ln P/d\ln k$, i.e., the integral is dominated by 
$k\sim 0.05 \ihmpc$, where $n_{\rm eff}\sim -1$.
I will leave a numerical evaluation of the detectability of the signal for the 
next section, where I include several other effects which change the
signal amplitude in detail, although not the order of magnitude or form.

\section{The redshift-space power spectrum to 
$\mathcal{O}(k_c^{-1})$ \label{secYoo}}

Recently, \cite{2009arXiv0907.0707Y} presented a much more rigorous and 
complete calculation of
the observable redshift-space galaxy density. Here I will use their result to 
compute the
redshift-space power spectrum, which amounts to changing the coefficients in 
$I^i_\vk$ above. When we start to include relativistic terms, it becomes
natural to look at the calculation as an expansion in $1/k_c$. I will compute
the first, $\mathcal{O}(k_c^{-1})$, correction to the standard LSS picture, 
dropping terms $\ordertwo$, 
which, as I discussed above, are very hard to detect (except in projection
against the very smooth CMB background 
\cite{2008PhRvD..78d3519H,2008PhRvD..77l3520G,2009ApJ...701..414G}).

Before considering galaxy clustering, it is useful to check if there are any
$k_c^{-1}$ corrections to the standard calculations of the perturbations in
mass-density, $\delta$, velocity, $v^i$,
and potential, $\psi$, fields. 
I will consider the evolution equations appropriate for
dark matter perturbations only, because we are interested in late times and 
large scales where baryons and dark matter are equivalent, and radiation is
negligible (I assume a 
cosmological constant, so no dark energy density fluctuations). 
I will present the results in terms of 
quantities calculated in the conformal Newtonian gauge, where the metric is:
\begin{equation}
ds^2 = g_{\mu\nu} dx^\mu dx^\nu =
a^2\left(\tau\right)\left[-\left(1+2 \psi\right)d\tau^2 +
\left(1+2 \phi\right)dx^2\right]~.
\end{equation}
$\psi$ can be identified with the Newtonian potential, and $\phi=-\psi$.
The final galaxy clustering results will be gauge invariant, because they are 
calculated entirely in terms of observable combinations of the gauge-dependent
quantities.
The evolution equations are \cite{2003moco.book.....D}:
\begin{equation}
\dot{\delta}+3 \dot{\phi}+\theta=0~,
\end{equation}
where $\theta = \partial_i v^i$ and the dot is a derivative with respect to 
conformal time
(note that this is the linearized Newtonian continuity equation plus an 
extra $\dot{\phi}$ term),
\begin{equation}
\dot{\theta}+\mathcal{H} \theta -k^2\psi=0
\end{equation}
where $\mathcal{H}=H a$
(just the standard linear Euler equation), and
\begin{equation}
k^2 \phi = 4 \pi G a^2 \bar{\rho}_m \left[\delta_m +
\frac{3 \mathcal{H}}{k^2}\theta\right]
\end{equation}
(the Poisson equation, plus the extra $\theta$ term).
We see immediately that all corrections are $\mathcal{O}(k_c^{-2})$, i.e., 
irrelevant to this calculation. We could have guessed this from symmetry -- 
all of the effects
at $\mathcal{O}(k_c^{-1})$ will be related to the observing process, where 
reflection symmetry is broken along the line of sight.

Most of the work needed to compute the galaxy power spectrum has been done
by \cite{2009arXiv0907.0707Y}. Their equation Eq. (36), which I write 
here in
terms of conformal Newtonian gauge sub-components, gives the gauge invariant
fluctuations in galaxy density, that one can observe using only the measurable
frequency of photons and angular position on the sky:
\begin{equation}
\label{eqyoo36}
\delta_{\rm obs} =b(\delta_m -3 \delta_z)+\psi+2 \phi +v_\parallel -
(1+z)\frac{\partial \delta_z}{\partial z}-2\frac{1+z}{H r}\delta_z-\delta_z
-5 p \delta_{\mathcal{D}_L}-2 \kappa+\frac{1+z}{H}\frac{dH}{dz}\delta_z+
2 \frac{\delta_r}{r}~,
\end{equation}
where the various
elements of this equation, discussed below, are derived in 
\cite{2009arXiv0907.0707Y}. 
I have not verified that this equation is correct. Note that it is certainly
not perfectly complete, e.g.,
evolution in the number density of galaxies would produce at least one more
term (this would not induce any interesting new behavior, so I continue to 
neglect it). The redshift $z$ is the observed one, and background quantities 
like $H$ are evaluated at this observed $z$. I am going to assume that the 
background quantities are known functions of $z$. Relaxing this assumption leads
to the usual Alcock-Paczynski-type effects \cite{1979Natur.281..358A}, 
which, it should be understood, are
different in nature from the ones that lead to Eq. (\ref{eqyoo36}). The
terms in Eq. (\ref{eqyoo36}) are related to the uncertainty in the $z$ at which
to evaluate the background functions (among other things), rather than 
uncertainty in the functions
themselves. One can perfectly well imagine studying
LSS in a Universe where the background evolution was known perfectly (e.g., 
for the purpose of studying the primordial perturbations), and in that case 
Eq. (\ref{eqyoo36}) would be the one to use.  

I now pare Eq. (\ref{eqyoo36}) down to $\mathcal{O}(k_c^{-1})$ terms, in 
addition to dropping the $\kappa$ (lensing) term for simplicity. Lensing can
be important \cite{2008PhRvD..77f3526H}, but the fact that it is sensitive to 
perturbations integrated 
along the line of sight means that it produces a power spectrum with a much 
different form than the terms that I consider, outside the main point 
of this paper. Everything I consider
here would be called primary fluctuations in CMB language, i.e., not depending
on an integral over perturbations along the line of sight.
Eq. (16) of \cite{2009arXiv0907.0707Y} shows that $\delta r/r \sim 
\mathcal{O}(k_c^{-2})$ so it can be dropped. Note that $r$ is the comoving
distance to the galaxy.  
The redshift perturbation $\delta_z$ can be inferred from Eq. (11) of
\cite{2009arXiv0907.0707Y}, and the subsequent paragraph:
\begin{equation}
\delta_z= v_\parallel-\psi-
\int_0^{r} dr^\prime \left(\dot{\psi}-\dot{\phi}\right)
\end{equation}
or
\begin{equation}
\delta_z= v_\parallel+\ordertwo
\end{equation}
and
\begin{equation}
\frac{\partial\delta_z}{\partial z}=
\frac{\partial v_\parallel}{\partial z}
-\frac{\partial \psi}{\partial z}+\ordertwo
\end{equation}
(Recall that $v\propto k^{-1} \delta$, $\psi\propto k^{-2} \delta$, and
a derivative brings in an extra factor of $k$.)

Eq. (27) of \cite{2009arXiv0907.0707Y} gives
\begin{equation}
\delta_{\mathcal{D}_L}= v_\parallel-\frac{1+z}{H r}\delta_z 
+\ordertwo
\end{equation}
where I have dropped the lensing component. 
Note that the luminosity bias term should be viewed as a place-holder for a 
variety of
possible similar terms that can appear, depending on the type of observable,
i.e., it assumes our observable is a simple observed-magnitude-limited galaxy
density, while in reality we may be using other things like halo-mass-weighted 
density \cite{2009arXiv0904.2963S} or 21cm intensity mapping 
\cite{2008PhRvL.100i1303C} or the \lyaf\ 
\cite{2007PhRvD..76f3009M,2006ApJS..163...80M}.
One might be tempted to make a ``distant observer'' approximation and drop the
term $\propto c/ \mathcal{H} r$ (note that I usually use $c=1$), however this 
is not a good approximation, e.g., at $z=1.5$,  
$c/ \mathcal{H} r \sim 1$. In the language of this paper, the usual
justification for dropping this kind of term is actually the $k_c^{-1}$ 
expansion, only breaking down if $1/r$ becomes large enough to overcome this.

Using $(1+z)\partial_z=\mathcal{H}^{-1}\partial_r$, we are left with
\begin{equation}
\delta_{\rm obs} =
b \delta_m  
-\frac{1}{\mathcal{H}}\frac{\partial v_\parallel}{\partial r}
+\frac{1}{\mathcal{H}}\frac{\partial \psi}{\partial r}
-\left(3 b+\frac{2}{\mathcal{H} r}+
5 p \left(1-\frac{1}{\mathcal{H} r}\right)-
\frac{1}{\mathcal{H}}\frac{dH}{dz}\right)v_\parallel
+\ordertwo+{\rm lensing}
\end{equation}
We see that the first two terms are the usual redshift-space density field, 
the third is the gravitational redshift term computed in \S\ref{secnaive}, and
then there are a variety of terms proportional to $v_\parallel$.

Moving to Fourier space, we see that the $v_\parallel$ term has the same $\vk$
dependence as the gravitational redshift term, i.e., 
$v^\parallel_\vk=-i f \frac{\mu}{ k_c} \delta_\vk$, while
$-i k_\parallel \psi_\vk/\mathcal{H}=
i\frac{3}{2}\Omega_m \frac{\mu}{k_c} \delta_\vk$, the difference
is only in the coefficients. Finally,
\begin{equation}
\delta_{\vk}^{\rm obs} =
\left[b+f\mu^2 +i\frac{\mu}{k_c}\left(\frac{3}{2}\Omega_m+
f\left\{3 b+\frac{2}{\mathcal{H} r}+5 p \left(1-\frac{1}{\mathcal{H} r}\right)+
\frac{d\ln H}{d\ln a}\right\}\right) \right]
\delta_\vk ~.
\end{equation}
Note that the Fourier transform in the radial direction involves a bit of a 
slight of hand, as the statistics of the field are not stationary 
(distance-independent) in that direction. This is not a new problem, as similar
redshift evolution has always been present in this kind of calculation within 
the growth factor (hidden in $\delta(z)$) and $f(z)$ and $b(z)$, but it is 
highlighted by the explicit presence of the comoving distance $r(z)$ itself 
(which is also the coordinate in the Fourier transform (it may be 
more comforting to think of $\mathcal{H}r$ as the Alcock-Paczynski factor)). 
Of course, the Fourier
transform in this type of calculation is also somewhat formal because the 
surveys are not infinite or periodic.
Dealing with these issues will require very careful data
analysis, presumably done at least partially in configuration space.
In the end, although they will make one's life 
more complicated, it is likely that background uncertainty and evolution issues
will lead 
to new opportunities to derive information from these surveys, as exemplified
by the Alcock-Paczynski test \cite{1979Natur.281..358A}. 

I now use Eq. (\ref{eqgeneralPI}) to compute the imaginary part of the
power spectrum for two types of galaxy,
\begin{eqnarray}
\frac{P_I^{ab}(k,\mu)}{P(k)}&=&\frac{\mu}{k_c}\left[\left(\frac{3}{2}\Omega_m +f
\left\{\frac{2}{\mathcal{H}r}+\frac{d\ln H}{d\ln a}\right\}\right)
\left(b_b-b_a\right)+5\left(1-\frac{1}{\mathcal{H}r}\right)\left(b_b p_a-
b_a p_b\right)\right] \nonumber \\
&+& f \frac{\mu^3}{k_c}\left[3 f \left(b_a-b_b\right)+
5\left(1-\frac{1}{\mathcal{H}r}\right)\left(p_a-p_b\right)\right] ~.  
\label{eqmainresult}
\end{eqnarray}
One thing to note about this equation is that
$\frac{3}{2}\Omega_m +f \frac{d\ln H}{d\ln a}=0$ in an Einstein-de Sitter 
Universe, and will therefore be fairly small in the real Universe, except at 
low $z$ (where there is not a lot of volume). $1-\frac{1}{\mathcal{H}r}$ is not
necessarily quite as generally small, but is nearly zero at $z\sim 1.5$ where 
a JDEM-like survey would be targeted. Beyond that, it is useful to plug in 
some numbers in order to do a concrete calculation of 
detectability.
I will assume $b_a=2$, $b_b=1$ (this level of difference is quite reasonable, 
e.g., the BigBOSS proposal \cite{2009arXiv0904.0468S}
contains two different samples with roughly these biases, 
although not observed on quite the scale assumed here), $p_a=1$, $p_b=0.5$
(roughly reasonable \cite{2009arXiv0907.0707Y,2008arXiv0807.3551G}, and not
very important), and $\Omega_{m,0}=0.28$ \cite{2006JCAP...10..014S}.
This gives $(\mathcal{H}r)^{-1}= 1.08$ at $z=1.5$, $\Omega_m(z=1.5)=0.86$, 
$f(z=1.5)=0.92$, and $d\ln H/d\ln a=-1.29$.
This makes 
$P_I^{ab}(k,\mu)=\left(\mu/k_c\right)\left(-2.1+2.74\mu^2\right)P(k)$. 
Finally, I assume a 100 cubic Gpc/h survey, and noise given by 
$\bar{n}^i P^i(k=0.2\ihmpc)=1$, where $\bar{n}^i=1/N^i$ is the number density
of galaxies of type $i$. This noise level is often used as a target for BAO
surveys, although $nP\sim 3$ is really required to squeeze out most of the 
BAO information.  
Integrating Eq. (\ref{eqSN}), I find 
$S/N=3$, when the integration is taken to $k=0.2\ihmpc$ ($S/N=2.7$ if the limit
is $0.1 \ihmpc$). To scale this to other surveys, $S/N\propto V^{1/2}$ and 
$S/N\propto \bar{n}^{1/2}$, i.e., 
we would reach a 5-$\sigma$ detection for $nP=3$.  

\section{Conclusions \label{secconclusions}}

The main result of this paper is Eq. (\ref{eqmainresult}), the imaginary part
of the cross-power spectrum between two different types of galaxies, which is
generated by gravitational redshift and other effects of similar form in the 
redshift-space density field. The
signal is suppressed by $\frac{a H}{c k}$ relative to the standard power 
spectrum, but the increasing number of modes at higher
$k$ makes the typical scale contributing to a detection relatively small, i.e.,
$k\sim 0.05 \ihmpc$.
This effect should be detectable in JDEM-scale \cite{2009arXiv0901.0721A}
surveys, although it does require subsets of galaxies with substantially 
different bias, both with good signal power-to-noise power ratios, so a 
bare-bones BAO experiment is not necessarily optimal -- a 100 cubic Gpc/h 
survey that can 
achieve $nP(k=0.2\ihmpc)=1$ 
for two samples of galaxies with a separation in biases of 
$\sim 1$ would make a $\sim 3-\sigma$ detection. 
The measurement errors will be dominated by shot-noise, not sample variance, so
there is a lot of room for improvement.
21cm intensity mapping \cite{2008PhRvL.100i1303C} could provide a perfect low
bias, high S/N field complementing a high bias BAO-oriented galaxy survey.
Creative halo weighting schemes \cite{2009arXiv0904.2963S} promise to change 
the way we think about shot-noise, possibly reducing the noise power in planned
surveys by an order of magnitude below $\bar{n}^{-1}$.

Taking the imaginary part of the power spectrum naturally has the effect of 
beating cosmic variance in the sense of 
\cite{2009PhRvL.102b1302S,2008arXiv0810.0323M}, so it isn't clear that anything
more can be gained in that direction.
This result continues the recent trend toward viewing the fact that different
kinds of galaxies have different biases, i.e., trace the density field in 
different ways, as an opportunity to tease out different physical effects, 
rather than simply as a nuisance 
\cite{2009arXiv0904.2963S,2009PhRvL.102b1302S,2008arXiv0810.0323M}.
While the short-term planning of surveys will probably continue to follow the
``measure the BAO feature by looking for high bias and volume with 
$nP\sim 1$'' strategy, it is clear that the long-term 
future of LSS studies is much richer than that. 


\acknowledgements

I thank Paul Shapiro for a question that led me to compute the gravitational
redshift effect, and   
Uro\v s Seljak for helpful comments. I acknowledge support
of the Beatrice D. Tremaine Fellowship.
 
\bibliography{$LATEX/cosmo,$LATEX/cosmo_preprints}

\begin{thebibliography}{49}
\expandafter\ifx\csname natexlab\endcsname\relax\def\natexlab#1{#1}\fi
\expandafter\ifx\csname bibnamefont\endcsname\relax
  \def\bibnamefont#1{#1}\fi
\expandafter\ifx\csname bibfnamefont\endcsname\relax
  \def\bibfnamefont#1{#1}\fi
\expandafter\ifx\csname citenamefont\endcsname\relax
  \def\citenamefont#1{#1}\fi
\expandafter\ifx\csname url\endcsname\relax
  \def\url#1{\texttt{#1}}\fi
\expandafter\ifx\csname urlprefix\endcsname\relax\def\urlprefix{URL }\fi
\providecommand{\bibinfo}[2]{#2}
\providecommand{\eprint}[2][]{\url{#2}}

\bibitem[{\citenamefont{{Kaiser}}(1987)}]{1987MNRAS.227....1K}
\bibinfo{author}{\bibfnamefont{N.}~\bibnamefont{{Kaiser}}},
  \bibinfo{journal}{\mnras} \textbf{\bibinfo{volume}{227}}, \bibinfo{pages}{1}
  (\bibinfo{year}{1987}).

\bibitem[{\citenamefont{{McDonald} and {Seljak}}(2008)}]{2008arXiv0810.0323M}
\bibinfo{author}{\bibfnamefont{P.}~\bibnamefont{{McDonald}}} \bibnamefont{and}
  \bibinfo{author}{\bibfnamefont{U.}~\bibnamefont{{Seljak}}},
  \bibinfo{journal}{ArXiv e-prints}  (\bibinfo{year}{2008}),
  \eprint{0810.0323}.

\bibitem[{\citenamefont{{White} et~al.}(2009)\citenamefont{{White}, {Song}, and
  {Percival}}}]{2009MNRAS.tmp..925W}
\bibinfo{author}{\bibfnamefont{M.}~\bibnamefont{{White}}},
  \bibinfo{author}{\bibfnamefont{Y.-S.} \bibnamefont{{Song}}},
  \bibnamefont{and} \bibinfo{author}{\bibfnamefont{W.~J.}
  \bibnamefont{{Percival}}}, \bibinfo{journal}{\mnras} pp.
  \bibinfo{pages}{925--+} (\bibinfo{year}{2009}), \eprint{0810.1518}.

\bibitem[{\citenamefont{{Zhang} et~al.}(2007)\citenamefont{{Zhang}, {Liguori},
  {Bean}, and {Dodelson}}}]{2007PhRvL..99n1302Z}
\bibinfo{author}{\bibfnamefont{P.}~\bibnamefont{{Zhang}}},
  \bibinfo{author}{\bibfnamefont{M.}~\bibnamefont{{Liguori}}},
  \bibinfo{author}{\bibfnamefont{R.}~\bibnamefont{{Bean}}}, \bibnamefont{and}
  \bibinfo{author}{\bibfnamefont{S.}~\bibnamefont{{Dodelson}}},
  \bibinfo{journal}{Physical Review Letters} \textbf{\bibinfo{volume}{99}},
  \bibinfo{pages}{141302} (\bibinfo{year}{2007}), \eprint{0704.1932}.

\bibitem[{\citenamefont{{Zhang}}(2008)}]{2008arXiv0802.2416Z}
\bibinfo{author}{\bibfnamefont{P.}~\bibnamefont{{Zhang}}},
  \bibinfo{journal}{ArXiv e-prints}  (\bibinfo{year}{2008}),
  \eprint{0802.2416}.

\bibitem[{\citenamefont{{Percival} and {White}}(2009)}]{2009MNRAS.393..297P}
\bibinfo{author}{\bibfnamefont{W.~J.} \bibnamefont{{Percival}}}
  \bibnamefont{and} \bibinfo{author}{\bibfnamefont{M.}~\bibnamefont{{White}}},
  \bibinfo{journal}{\mnras} \textbf{\bibinfo{volume}{393}},
  \bibinfo{pages}{297} (\bibinfo{year}{2009}), \eprint{0808.0003}.

\bibitem[{\citenamefont{{Taruya} et~al.}(2009)\citenamefont{{Taruya},
  {Nishimichi}, {Saito}, and {Hiramatsu}}}]{2009arXiv0906.0507T}
\bibinfo{author}{\bibfnamefont{A.}~\bibnamefont{{Taruya}}},
  \bibinfo{author}{\bibfnamefont{T.}~\bibnamefont{{Nishimichi}}},
  \bibinfo{author}{\bibfnamefont{S.}~\bibnamefont{{Saito}}}, \bibnamefont{and}
  \bibinfo{author}{\bibfnamefont{T.}~\bibnamefont{{Hiramatsu}}},
  \bibinfo{journal}{ArXiv e-prints}  (\bibinfo{year}{2009}),
  \eprint{0906.0507}.

\bibitem[{\citenamefont{{Shaw} and {Lewis}}(2008)}]{2008PhRvD..78j3512S}
\bibinfo{author}{\bibfnamefont{J.~R.} \bibnamefont{{Shaw}}} \bibnamefont{and}
  \bibinfo{author}{\bibfnamefont{A.}~\bibnamefont{{Lewis}}},
  \bibinfo{journal}{\prd} \textbf{\bibinfo{volume}{78}},
  \bibinfo{pages}{103512} (\bibinfo{year}{2008}), \eprint{0808.1724}.

\bibitem[{\citenamefont{{Scoccimarro}}(2004)}]{2004PhRvD..70h3007S}
\bibinfo{author}{\bibfnamefont{R.}~\bibnamefont{{Scoccimarro}}},
  \bibinfo{journal}{\prd} \textbf{\bibinfo{volume}{70}},
  \bibinfo{pages}{083007} (\bibinfo{year}{2004}).

\bibitem[{\citenamefont{{Matsubara}}(2008{\natexlab{a}})}]{2008PhRvD..77f3530M}
\bibinfo{author}{\bibfnamefont{T.}~\bibnamefont{{Matsubara}}},
  \bibinfo{journal}{\prd} \textbf{\bibinfo{volume}{77}},
  \bibinfo{pages}{063530} (\bibinfo{year}{2008}{\natexlab{a}}),
  \eprint{arXiv:0711.2521}.

\bibitem[{\citenamefont{{Matsubara}}(2008{\natexlab{b}})}]{2008PhRvD..78h3519M}
\bibinfo{author}{\bibfnamefont{T.}~\bibnamefont{{Matsubara}}},
  \bibinfo{journal}{\prd} \textbf{\bibinfo{volume}{78}},
  \bibinfo{pages}{083519} (\bibinfo{year}{2008}{\natexlab{b}}).

\bibitem[{\citenamefont{{Nishimichi} et~al.}(2007)\citenamefont{{Nishimichi},
  {Ohmuro}, {Nakamichi}, {Taruya}, {Yahata}, {Shirata}, {Saito}, {Nomura},
  {Yamamoto}, and {Suto}}}]{2007PASJ...59.1049N}
\bibinfo{author}{\bibfnamefont{T.}~\bibnamefont{{Nishimichi}}},
  \bibinfo{author}{\bibfnamefont{H.}~\bibnamefont{{Ohmuro}}},
  \bibinfo{author}{\bibfnamefont{M.}~\bibnamefont{{Nakamichi}}},
  \bibinfo{author}{\bibfnamefont{A.}~\bibnamefont{{Taruya}}},
  \bibinfo{author}{\bibfnamefont{K.}~\bibnamefont{{Yahata}}},
  \bibinfo{author}{\bibfnamefont{A.}~\bibnamefont{{Shirata}}},
  \bibinfo{author}{\bibfnamefont{S.}~\bibnamefont{{Saito}}},
  \bibinfo{author}{\bibfnamefont{H.}~\bibnamefont{{Nomura}}},
  \bibinfo{author}{\bibfnamefont{K.}~\bibnamefont{{Yamamoto}}},
  \bibnamefont{and} \bibinfo{author}{\bibfnamefont{Y.}~\bibnamefont{{Suto}}},
  \bibinfo{journal}{\pasj} \textbf{\bibinfo{volume}{59}}, \bibinfo{pages}{1049}
  (\bibinfo{year}{2007}), \eprint{0705.1589}.

\bibitem[{\citenamefont{{Heavens} et~al.}(1998)\citenamefont{{Heavens},
  {Matarrese}, and {Verde}}}]{1998MNRAS.301..797H}
\bibinfo{author}{\bibfnamefont{A.~F.} \bibnamefont{{Heavens}}},
  \bibinfo{author}{\bibfnamefont{S.}~\bibnamefont{{Matarrese}}},
  \bibnamefont{and} \bibinfo{author}{\bibfnamefont{L.}~\bibnamefont{{Verde}}},
  \bibinfo{journal}{\mnras} \textbf{\bibinfo{volume}{301}},
  \bibinfo{pages}{797} (\bibinfo{year}{1998}).

\bibitem[{\citenamefont{{McDonald} and {Roy}}(2009)}]{2009arXiv0902.0991M}
\bibinfo{author}{\bibfnamefont{P.}~\bibnamefont{{McDonald}}} \bibnamefont{and}
  \bibinfo{author}{\bibfnamefont{A.}~\bibnamefont{{Roy}}},
  \bibinfo{journal}{ArXiv e-prints}  (\bibinfo{year}{2009}),
  \eprint{0902.0991}.

\bibitem[{\citenamefont{{McDonald} et~al.}(2000)\citenamefont{{McDonald},
  {Miralda-Escud{\' e}}, {Rauch}, {Sargent}, {Barlow}, {Cen}, and
  {Ostriker}}}]{2000ApJ...543....1M}
\bibinfo{author}{\bibfnamefont{P.}~\bibnamefont{{McDonald}}},
  \bibinfo{author}{\bibfnamefont{J.}~\bibnamefont{{Miralda-Escud{\' e}}}},
  \bibinfo{author}{\bibfnamefont{M.}~\bibnamefont{{Rauch}}},
  \bibinfo{author}{\bibfnamefont{W.~L.~W.} \bibnamefont{{Sargent}}},
  \bibinfo{author}{\bibfnamefont{T.~A.} \bibnamefont{{Barlow}}},
  \bibinfo{author}{\bibfnamefont{R.}~\bibnamefont{{Cen}}}, \bibnamefont{and}
  \bibinfo{author}{\bibfnamefont{J.~P.} \bibnamefont{{Ostriker}}},
  \bibinfo{journal}{\apj} \textbf{\bibinfo{volume}{543}}, \bibinfo{pages}{1}
  (\bibinfo{year}{2000}).

\bibitem[{\citenamefont{{McDonald}}(2003)}]{2003ApJ...585...34M}
\bibinfo{author}{\bibfnamefont{P.}~\bibnamefont{{McDonald}}},
  \bibinfo{journal}{\apj} \textbf{\bibinfo{volume}{585}}, \bibinfo{pages}{34}
  (\bibinfo{year}{2003}), \eprint{arXiv:astro-ph/0108064}.

\bibitem[{\citenamefont{{Hirata}}(2009)}]{2009arXiv0903.4929H}
\bibinfo{author}{\bibfnamefont{C.~M.} \bibnamefont{{Hirata}}},
  \bibinfo{journal}{ArXiv e-prints}  (\bibinfo{year}{2009}),
  \eprint{0903.4929}.

\bibitem[{\citenamefont{{Yoo}}(2009)}]{2009PhRvD..79b3517Y}
\bibinfo{author}{\bibfnamefont{J.}~\bibnamefont{{Yoo}}},
  \bibinfo{journal}{\prd} \textbf{\bibinfo{volume}{79}},
  \bibinfo{pages}{023517} (\bibinfo{year}{2009}), \eprint{0808.3138}.

\bibitem[{\citenamefont{{Yoo} et~al.}(2009)\citenamefont{{Yoo}, {Fitzpatrick},
  and {Zaldarriaga}}}]{2009arXiv0907.0707Y}
\bibinfo{author}{\bibfnamefont{J.}~\bibnamefont{{Yoo}}},
  \bibinfo{author}{\bibfnamefont{A.~L.} \bibnamefont{{Fitzpatrick}}},
  \bibnamefont{and}
  \bibinfo{author}{\bibfnamefont{M.}~\bibnamefont{{Zaldarriaga}}},
  \bibinfo{journal}{ArXiv e-prints}  (\bibinfo{year}{2009}),
  \eprint{0907.0707}.

\bibitem[{\citenamefont{{P{\'a}pai} and {Szapudi}}(2008)}]{2008MNRAS.389..292P}
\bibinfo{author}{\bibfnamefont{P.}~\bibnamefont{{P{\'a}pai}}} \bibnamefont{and}
  \bibinfo{author}{\bibfnamefont{I.}~\bibnamefont{{Szapudi}}},
  \bibinfo{journal}{\mnras} \textbf{\bibinfo{volume}{389}},
  \bibinfo{pages}{292} (\bibinfo{year}{2008}), \eprint{0802.2940}.

\bibitem[{\citenamefont{{Albrecht} et~al.}(2009)\citenamefont{{Albrecht},
  {Amendola}, {Bernstein}, {Clowe}, {Eisenstein}, {Guzzo}, {Hirata}, {Huterer},
  {Kirshner}, {Kolb} et~al.}}]{2009arXiv0901.0721A}
\bibinfo{author}{\bibfnamefont{A.}~\bibnamefont{{Albrecht}}},
  \bibinfo{author}{\bibfnamefont{L.}~\bibnamefont{{Amendola}}},
  \bibinfo{author}{\bibfnamefont{G.}~\bibnamefont{{Bernstein}}},
  \bibinfo{author}{\bibfnamefont{D.}~\bibnamefont{{Clowe}}},
  \bibinfo{author}{\bibfnamefont{D.}~\bibnamefont{{Eisenstein}}},
  \bibinfo{author}{\bibfnamefont{L.}~\bibnamefont{{Guzzo}}},
  \bibinfo{author}{\bibfnamefont{C.}~\bibnamefont{{Hirata}}},
  \bibinfo{author}{\bibfnamefont{D.}~\bibnamefont{{Huterer}}},
  \bibinfo{author}{\bibfnamefont{R.}~\bibnamefont{{Kirshner}}},
  \bibinfo{author}{\bibfnamefont{E.}~\bibnamefont{{Kolb}}},
  \bibnamefont{et~al.}, \bibinfo{journal}{ArXiv e-prints}
  (\bibinfo{year}{2009}), \eprint{0901.0721}.

\bibitem[{\citenamefont{{McDonald}}(2008)}]{2008PhRvD..78l3519M}
\bibinfo{author}{\bibfnamefont{P.}~\bibnamefont{{McDonald}}},
  \bibinfo{journal}{\prd} \textbf{\bibinfo{volume}{78}},
  \bibinfo{pages}{123519} (\bibinfo{year}{2008}), \eprint{0806.1061}.

\bibitem[{\citenamefont{{Dalal} et~al.}(2008)\citenamefont{{Dalal}, {Dor{\'e}},
  {Huterer}, and {Shirokov}}}]{2008PhRvD..77l3514D}
\bibinfo{author}{\bibfnamefont{N.}~\bibnamefont{{Dalal}}},
  \bibinfo{author}{\bibfnamefont{O.}~\bibnamefont{{Dor{\'e}}}},
  \bibinfo{author}{\bibfnamefont{D.}~\bibnamefont{{Huterer}}},
  \bibnamefont{and}
  \bibinfo{author}{\bibfnamefont{A.}~\bibnamefont{{Shirokov}}},
  \bibinfo{journal}{\prd} \textbf{\bibinfo{volume}{77}},
  \bibinfo{pages}{123514} (\bibinfo{year}{2008}), \eprint{0710.4560}.

\bibitem[{\citenamefont{{Desjacques} and {Seljak}}(2009)}]{2009arXiv0907.2257D}
\bibinfo{author}{\bibfnamefont{V.}~\bibnamefont{{Desjacques}}}
  \bibnamefont{and} \bibinfo{author}{\bibfnamefont{U.}~\bibnamefont{{Seljak}}},
  \bibinfo{journal}{ArXiv e-prints}  (\bibinfo{year}{2009}),
  \eprint{0907.2257}.

\bibitem[{\citenamefont{{Desjacques} et~al.}(2009)\citenamefont{{Desjacques},
  {Seljak}, and {Iliev}}}]{2009MNRAS.396...85D}
\bibinfo{author}{\bibfnamefont{V.}~\bibnamefont{{Desjacques}}},
  \bibinfo{author}{\bibfnamefont{U.}~\bibnamefont{{Seljak}}}, \bibnamefont{and}
  \bibinfo{author}{\bibfnamefont{I.~T.} \bibnamefont{{Iliev}}},
  \bibinfo{journal}{\mnras} \textbf{\bibinfo{volume}{396}}, \bibinfo{pages}{85}
  (\bibinfo{year}{2009}), \eprint{0811.2748}.

\bibitem[{\citenamefont{{Sefusatti}}(2009)}]{2009arXiv0905.0717S}
\bibinfo{author}{\bibfnamefont{E.}~\bibnamefont{{Sefusatti}}},
  \bibinfo{journal}{ArXiv e-prints}  (\bibinfo{year}{2009}),
  \eprint{0905.0717}.

\bibitem[{\citenamefont{{Gong} et~al.}(2009)\citenamefont{{Gong}, {Wang},
  {Zheng}, and {Chen}}}]{2009arXiv0904.4257G}
\bibinfo{author}{\bibfnamefont{Y.}~\bibnamefont{{Gong}}},
  \bibinfo{author}{\bibfnamefont{X.}~\bibnamefont{{Wang}}},
  \bibinfo{author}{\bibfnamefont{Z.}~\bibnamefont{{Zheng}}}, \bibnamefont{and}
  \bibinfo{author}{\bibfnamefont{X.}~\bibnamefont{{Chen}}},
  \bibinfo{journal}{ArXiv e-prints}  (\bibinfo{year}{2009}),
  \eprint{0904.4257}.

\bibitem[{\citenamefont{{Crociani} et~al.}(2009)\citenamefont{{Crociani},
  {Moscardini}, {Viel}, and {Matarrese}}}]{2009MNRAS.394..133C}
\bibinfo{author}{\bibfnamefont{D.}~\bibnamefont{{Crociani}}},
  \bibinfo{author}{\bibfnamefont{L.}~\bibnamefont{{Moscardini}}},
  \bibinfo{author}{\bibfnamefont{M.}~\bibnamefont{{Viel}}}, \bibnamefont{and}
  \bibinfo{author}{\bibfnamefont{S.}~\bibnamefont{{Matarrese}}},
  \bibinfo{journal}{\mnras} \textbf{\bibinfo{volume}{394}},
  \bibinfo{pages}{133} (\bibinfo{year}{2009}), \eprint{0809.3909}.

\bibitem[{\citenamefont{{Jeong} and {Komatsu}}(2009)}]{2009arXiv0904.0497J}
\bibinfo{author}{\bibfnamefont{D.}~\bibnamefont{{Jeong}}} \bibnamefont{and}
  \bibinfo{author}{\bibfnamefont{E.}~\bibnamefont{{Komatsu}}},
  \bibinfo{journal}{ArXiv e-prints}  (\bibinfo{year}{2009}),
  \eprint{0904.0497}.

\bibitem[{\citenamefont{{Slosar}}(2009)}]{2009JCAP...03..004S}
\bibinfo{author}{\bibfnamefont{A.}~\bibnamefont{{Slosar}}},
  \bibinfo{journal}{Journal of Cosmology and Astro-Particle Physics}
  \textbf{\bibinfo{volume}{3}}, \bibinfo{pages}{4} (\bibinfo{year}{2009}),
  \eprint{0808.0044}.

\bibitem[{\citenamefont{{Grossi} et~al.}(2009)\citenamefont{{Grossi}, {Verde},
  {Carbone}, {Dolag}, {Branchini}, {Iannuzzi}, {Matarrese}, and
  {Moscardini}}}]{2009arXiv0902.2013G}
\bibinfo{author}{\bibfnamefont{M.}~\bibnamefont{{Grossi}}},
  \bibinfo{author}{\bibfnamefont{L.}~\bibnamefont{{Verde}}},
  \bibinfo{author}{\bibfnamefont{C.}~\bibnamefont{{Carbone}}},
  \bibinfo{author}{\bibfnamefont{K.}~\bibnamefont{{Dolag}}},
  \bibinfo{author}{\bibfnamefont{E.}~\bibnamefont{{Branchini}}},
  \bibinfo{author}{\bibfnamefont{F.}~\bibnamefont{{Iannuzzi}}},
  \bibinfo{author}{\bibfnamefont{S.}~\bibnamefont{{Matarrese}}},
  \bibnamefont{and}
  \bibinfo{author}{\bibfnamefont{L.}~\bibnamefont{{Moscardini}}},
  \bibinfo{journal}{ArXiv e-prints}  (\bibinfo{year}{2009}),
  \eprint{0902.2013}.

\bibitem[{\citenamefont{{Pillepich} et~al.}(2008)\citenamefont{{Pillepich},
  {Porciani}, and {Hahn}}}]{2008arXiv0811.4176P}
\bibinfo{author}{\bibfnamefont{A.}~\bibnamefont{{Pillepich}}},
  \bibinfo{author}{\bibfnamefont{C.}~\bibnamefont{{Porciani}}},
  \bibnamefont{and} \bibinfo{author}{\bibfnamefont{O.}~\bibnamefont{{Hahn}}},
  \bibinfo{journal}{ArXiv e-prints}  (\bibinfo{year}{2008}),
  \eprint{0811.4176}.

\bibitem[{\citenamefont{{Taruya} et~al.}(2008)\citenamefont{{Taruya}, {Koyama},
  and {Matsubara}}}]{2008PhRvD..78l3534T}
\bibinfo{author}{\bibfnamefont{A.}~\bibnamefont{{Taruya}}},
  \bibinfo{author}{\bibfnamefont{K.}~\bibnamefont{{Koyama}}}, \bibnamefont{and}
  \bibinfo{author}{\bibfnamefont{T.}~\bibnamefont{{Matsubara}}},
  \bibinfo{journal}{\prd} \textbf{\bibinfo{volume}{78}},
  \bibinfo{pages}{123534} (\bibinfo{year}{2008}), \eprint{0808.4085}.

\bibitem[{\citenamefont{{Afshordi} and {Tolley}}(2008)}]{2008PhRvD..78l3507A}
\bibinfo{author}{\bibfnamefont{N.}~\bibnamefont{{Afshordi}}} \bibnamefont{and}
  \bibinfo{author}{\bibfnamefont{A.~J.} \bibnamefont{{Tolley}}},
  \bibinfo{journal}{\prd} \textbf{\bibinfo{volume}{78}},
  \bibinfo{pages}{123507} (\bibinfo{year}{2008}), \eprint{0806.1046}.

\bibitem[{\citenamefont{{Carbone} et~al.}(2008)\citenamefont{{Carbone},
  {Verde}, and {Matarrese}}}]{2008ApJ...684L...1C}
\bibinfo{author}{\bibfnamefont{C.}~\bibnamefont{{Carbone}}},
  \bibinfo{author}{\bibfnamefont{L.}~\bibnamefont{{Verde}}}, \bibnamefont{and}
  \bibinfo{author}{\bibfnamefont{S.}~\bibnamefont{{Matarrese}}},
  \bibinfo{journal}{\apjl} \textbf{\bibinfo{volume}{684}}, \bibinfo{pages}{L1}
  (\bibinfo{year}{2008}), \eprint{0806.1950}.

\bibitem[{\citenamefont{{Granett} et~al.}(2009)\citenamefont{{Granett},
  {Neyrinck}, and {Szapudi}}}]{2009ApJ...701..414G}
\bibinfo{author}{\bibfnamefont{B.~R.} \bibnamefont{{Granett}}},
  \bibinfo{author}{\bibfnamefont{M.~C.} \bibnamefont{{Neyrinck}}},
  \bibnamefont{and}
  \bibinfo{author}{\bibfnamefont{I.}~\bibnamefont{{Szapudi}}},
  \bibinfo{journal}{\apj} \textbf{\bibinfo{volume}{701}}, \bibinfo{pages}{414}
  (\bibinfo{year}{2009}), \eprint{0812.1025}.

\bibitem[{\citenamefont{{Ho} et~al.}(2008)\citenamefont{{Ho}, {Hirata},
  {Padmanabhan}, {Seljak}, and {Bahcall}}}]{2008PhRvD..78d3519H}
\bibinfo{author}{\bibfnamefont{S.}~\bibnamefont{{Ho}}},
  \bibinfo{author}{\bibfnamefont{C.}~\bibnamefont{{Hirata}}},
  \bibinfo{author}{\bibfnamefont{N.}~\bibnamefont{{Padmanabhan}}},
  \bibinfo{author}{\bibfnamefont{U.}~\bibnamefont{{Seljak}}}, \bibnamefont{and}
  \bibinfo{author}{\bibfnamefont{N.}~\bibnamefont{{Bahcall}}},
  \bibinfo{journal}{\prd} \textbf{\bibinfo{volume}{78}},
  \bibinfo{pages}{043519} (\bibinfo{year}{2008}), \eprint{0801.0642}.

\bibitem[{\citenamefont{{Giannantonio}
  et~al.}(2008)\citenamefont{{Giannantonio}, {Scranton}, {Crittenden},
  {Nichol}, {Boughn}, {Myers}, and {Richards}}}]{2008PhRvD..77l3520G}
\bibinfo{author}{\bibfnamefont{T.}~\bibnamefont{{Giannantonio}}},
  \bibinfo{author}{\bibfnamefont{R.}~\bibnamefont{{Scranton}}},
  \bibinfo{author}{\bibfnamefont{R.~G.} \bibnamefont{{Crittenden}}},
  \bibinfo{author}{\bibfnamefont{R.~C.} \bibnamefont{{Nichol}}},
  \bibinfo{author}{\bibfnamefont{S.~P.} \bibnamefont{{Boughn}}},
  \bibinfo{author}{\bibfnamefont{A.~D.} \bibnamefont{{Myers}}},
  \bibnamefont{and} \bibinfo{author}{\bibfnamefont{G.~T.}
  \bibnamefont{{Richards}}}, \bibinfo{journal}{\prd}
  \textbf{\bibinfo{volume}{77}}, \bibinfo{pages}{123520}
  (\bibinfo{year}{2008}), \eprint{0801.4380}.

\bibitem[{\citenamefont{{Dodelson}}(2003)}]{2003moco.book.....D}
\bibinfo{author}{\bibfnamefont{S.}~\bibnamefont{{Dodelson}}},
  \emph{\bibinfo{title}{{Modern cosmology}}} (\bibinfo{year}{2003}).

\bibitem[{\citenamefont{{Alcock} and {Paczynski}}(1979)}]{1979Natur.281..358A}
\bibinfo{author}{\bibfnamefont{C.}~\bibnamefont{{Alcock}}} \bibnamefont{and}
  \bibinfo{author}{\bibfnamefont{B.}~\bibnamefont{{Paczynski}}},
  \bibinfo{journal}{\nat} \textbf{\bibinfo{volume}{281}}, \bibinfo{pages}{358}
  (\bibinfo{year}{1979}).

\bibitem[{\citenamefont{{Hui} et~al.}(2008)\citenamefont{{Hui},
  {Gazta{\~n}aga}, and {Loverde}}}]{2008PhRvD..77f3526H}
\bibinfo{author}{\bibfnamefont{L.}~\bibnamefont{{Hui}}},
  \bibinfo{author}{\bibfnamefont{E.}~\bibnamefont{{Gazta{\~n}aga}}},
  \bibnamefont{and}
  \bibinfo{author}{\bibfnamefont{M.}~\bibnamefont{{Loverde}}},
  \bibinfo{journal}{\prd} \textbf{\bibinfo{volume}{77}},
  \bibinfo{pages}{063526} (\bibinfo{year}{2008}), \eprint{0710.4191}.

\bibitem[{\citenamefont{{Seljak} et~al.}(2009)\citenamefont{{Seljak}, {Hamaus},
  and {Desjacques}}}]{2009arXiv0904.2963S}
\bibinfo{author}{\bibfnamefont{U.}~\bibnamefont{{Seljak}}},
  \bibinfo{author}{\bibfnamefont{N.}~\bibnamefont{{Hamaus}}}, \bibnamefont{and}
  \bibinfo{author}{\bibfnamefont{V.}~\bibnamefont{{Desjacques}}},
  \bibinfo{journal}{ArXiv e-prints}  (\bibinfo{year}{2009}),
  \eprint{0904.2963}.

\bibitem[{\citenamefont{{Chang} et~al.}(2008)\citenamefont{{Chang}, {Pen},
  {Peterson}, and {McDonald}}}]{2008PhRvL.100i1303C}
\bibinfo{author}{\bibfnamefont{T.-C.} \bibnamefont{{Chang}}},
  \bibinfo{author}{\bibfnamefont{U.-L.} \bibnamefont{{Pen}}},
  \bibinfo{author}{\bibfnamefont{J.~B.} \bibnamefont{{Peterson}}},
  \bibnamefont{and}
  \bibinfo{author}{\bibfnamefont{P.}~\bibnamefont{{McDonald}}},
  \bibinfo{journal}{Physical Review Letters} \textbf{\bibinfo{volume}{100}},
  \bibinfo{pages}{091303} (\bibinfo{year}{2008}), \eprint{0709.3672}.

\bibitem[{\citenamefont{{McDonald} and
  {Eisenstein}}(2007)}]{2007PhRvD..76f3009M}
\bibinfo{author}{\bibfnamefont{P.}~\bibnamefont{{McDonald}}} \bibnamefont{and}
  \bibinfo{author}{\bibfnamefont{D.~J.} \bibnamefont{{Eisenstein}}},
  \bibinfo{journal}{\prd} \textbf{\bibinfo{volume}{76}},
  \bibinfo{pages}{063009} (\bibinfo{year}{2007}),
  \eprint{arXiv:astro-ph/0607122}.

\bibitem[{\citenamefont{{McDonald} et~al.}(2006)\citenamefont{{McDonald},
  {Seljak}, {Burles}, {Schlegel}, {Weinberg}, {Cen}, {Shih}, {Schaye},
  {Schneider}, {Bahcall} et~al.}}]{2006ApJS..163...80M}
\bibinfo{author}{\bibfnamefont{P.}~\bibnamefont{{McDonald}}},
  \bibinfo{author}{\bibfnamefont{U.}~\bibnamefont{{Seljak}}},
  \bibinfo{author}{\bibfnamefont{S.}~\bibnamefont{{Burles}}},
  \bibinfo{author}{\bibfnamefont{D.~J.} \bibnamefont{{Schlegel}}},
  \bibinfo{author}{\bibfnamefont{D.~H.} \bibnamefont{{Weinberg}}},
  \bibinfo{author}{\bibfnamefont{R.}~\bibnamefont{{Cen}}},
  \bibinfo{author}{\bibfnamefont{D.}~\bibnamefont{{Shih}}},
  \bibinfo{author}{\bibfnamefont{J.}~\bibnamefont{{Schaye}}},
  \bibinfo{author}{\bibfnamefont{D.~P.} \bibnamefont{{Schneider}}},
  \bibinfo{author}{\bibfnamefont{N.~A.} \bibnamefont{{Bahcall}}},
  \bibnamefont{et~al.}, \bibinfo{journal}{\apjs}
  \textbf{\bibinfo{volume}{163}}, \bibinfo{pages}{80} (\bibinfo{year}{2006}),
  \eprint{arXiv:astro-ph/0405013}.

\bibitem[{\citenamefont{{Schlegel} et~al.}(2009)\citenamefont{{Schlegel},
  {Bebek}, {Heetderks}, {Ho}, {Lampton}, {Levi}, {Mostek}, {Padmanabhan},
  {Perlmutter}, {Roe} et~al.}}]{2009arXiv0904.0468S}
\bibinfo{author}{\bibfnamefont{D.~J.} \bibnamefont{{Schlegel}}},
  \bibinfo{author}{\bibfnamefont{C.}~\bibnamefont{{Bebek}}},
  \bibinfo{author}{\bibfnamefont{H.}~\bibnamefont{{Heetderks}}},
  \bibinfo{author}{\bibfnamefont{S.}~\bibnamefont{{Ho}}},
  \bibinfo{author}{\bibfnamefont{M.}~\bibnamefont{{Lampton}}},
  \bibinfo{author}{\bibfnamefont{M.}~\bibnamefont{{Levi}}},
  \bibinfo{author}{\bibfnamefont{N.}~\bibnamefont{{Mostek}}},
  \bibinfo{author}{\bibfnamefont{N.}~\bibnamefont{{Padmanabhan}}},
  \bibinfo{author}{\bibfnamefont{S.}~\bibnamefont{{Perlmutter}}},
  \bibinfo{author}{\bibfnamefont{N.}~\bibnamefont{{Roe}}},
  \bibnamefont{et~al.}, \bibinfo{journal}{ArXiv e-prints}
  (\bibinfo{year}{2009}), \eprint{0904.0468}.

\bibitem[{\citenamefont{{Gaztanaga} et~al.}(2008)\citenamefont{{Gaztanaga},
  {Cabre}, and {Hui}}}]{2008arXiv0807.3551G}
\bibinfo{author}{\bibfnamefont{E.}~\bibnamefont{{Gaztanaga}}},
  \bibinfo{author}{\bibfnamefont{A.}~\bibnamefont{{Cabre}}}, \bibnamefont{and}
  \bibinfo{author}{\bibfnamefont{L.}~\bibnamefont{{Hui}}},
  \bibinfo{journal}{ArXiv e-prints}  (\bibinfo{year}{2008}),
  \eprint{0807.3551}.

\bibitem[{\citenamefont{{Seljak} et~al.}(2006)\citenamefont{{Seljak}, {Slosar},
  and {McDonald}}}]{2006JCAP...10..014S}
\bibinfo{author}{\bibfnamefont{U.}~\bibnamefont{{Seljak}}},
  \bibinfo{author}{\bibfnamefont{A.}~\bibnamefont{{Slosar}}}, \bibnamefont{and}
  \bibinfo{author}{\bibfnamefont{P.}~\bibnamefont{{McDonald}}},
  \bibinfo{journal}{Journal of Cosmology and Astro-Particle Physics}
  \textbf{\bibinfo{volume}{10}}, \bibinfo{pages}{14} (\bibinfo{year}{2006}).

\bibitem[{\citenamefont{{Seljak}}(2009)}]{2009PhRvL.102b1302S}
\bibinfo{author}{\bibfnamefont{U.}~\bibnamefont{{Seljak}}},
  \bibinfo{journal}{Physical Review Letters} \textbf{\bibinfo{volume}{102}},
  \bibinfo{pages}{021302} (\bibinfo{year}{2009}), \eprint{0807.1770}.

\end{thebibliography}

\end{document}